\documentclass[multphys,vecphys]{svmult}

\usepackage{makeidx}         
\usepackage{graphicx}        
\usepackage{multicol}        

\makeindex             

\begin{document}
\title*{Clumps of material orbiting a black hole and the QPOs}
\author{Uro\v s Kosti\' c \and
Andrej \v Cade\v z
\and Andreja Gomboc}
\institute{Department of Physics, Faculty of Mathematics and Physics, University of Ljubljana,
Jadranska 19, 1000 Ljubljana
\texttt{andrej.cadez@fmf.uni-lj.si}}
\maketitle
\section{Successive passages of an asteroid about a
black hole}
Clumps of material, the size of asteroids, may orbit a black hole and be distributed, somewhat like comets in the Kuiper belt. Some clumps are perturbed to
relatively small periastron high eccentricity orbits. During each periastron passage tides do work and so change orbital 
parameters to make the orbit more eccentric with less orbital angular momentum and lower periastron.
The periastron crossing time decreases
accordingly. When the periastron touches the Roche
radius, tides resonate with the fundamental quadrupole
mode of the asteroid \cite{kos:2005ApJ...625..278G}. The
transfer of orbital angular momentum to internal tidal
modes is high and the work done by tides is large enough
to heat the asteroid to high temperatures and to break
loose some parts of the asteroid. Each next periastron passage has a shorter characteristic time and excites higher 
mechanical modes of the asteroid. The orbital energy keeps decreasing and the transferred energy
is used to heat it to higher and higher temperatures and to accelerate
more breaking-away parts. Soon the asteroid is
split into small pieces which are 
sufficiently small that they are 
stable against tidal destruction.
When the periastron reaches close to the last circular
orbit, the overwhelming tidal force crushes all solids that are larger than about $0.01 GM/c^2$
(here $M$ is the mass of the black hole; $GM_\odot/c^2= 1.5~$km) and produces enough work to heat the 
remnants to
hard X-ray temperatures. 

The overwhelming tidal force is very rapid
with respect to the frequency of fundamental modes of
the asteroid, so that the work done by tides can be as
high as $0.1 \mathrm{mc^2}$, where $\mathrm{m}$ is the mass of the asteroid. Even a small part of this energy is
more than enough to make the asteroid explode. Yet,
the shock wave is still slow with respect to asteroid's
relativistic speed, so that it has no time to grow large
with respect to the black hole. 
Since the work is taken out of 
orbital energy, the remnants can no longer return to
the far apastron, but are caught by the black hole and after a few turns about the black hole at the last circular orbit, they disappear behind its horizon.
In making these few last
turns they alternately move towards and away
from the observer close to asteroid's orbital plane, so that the Doppler effect, aberration
of light and possibly gravitational lensing make characteristic peaks (a chirp) in the light curve of this last
orbit as it is seen by the observer.

\section{Fourier spectra of orbiting clumps and QPOs}
To investigate what kind of the observed light curve one would obtain from infalling debris of 
tidally disrupted asteroid, 
we assume that a nonrotating black hole is fed by a large number of small clumps and
calculate light curves from such events and their Fourier spectra.
\begin{figure}
\centering
\includegraphics[width=0.49\textwidth]{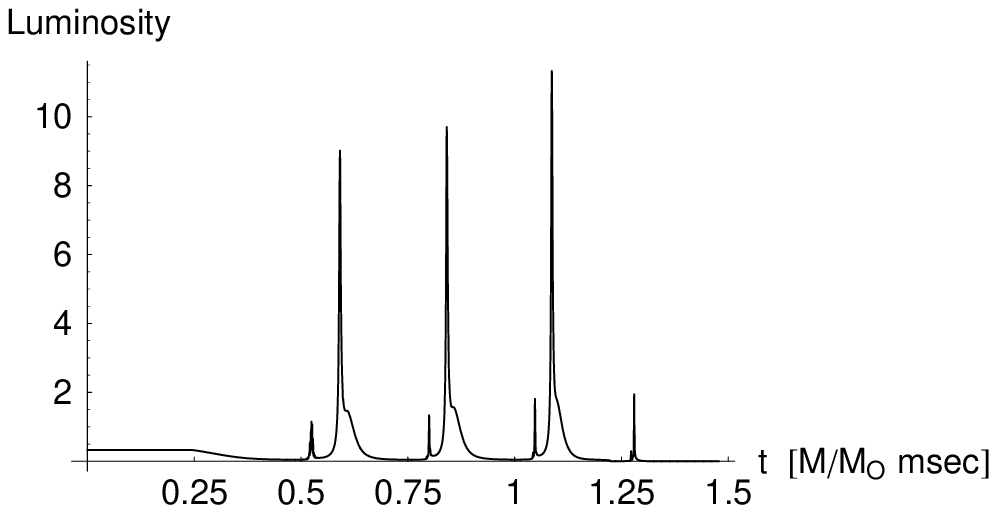}
\includegraphics[width=0.49\textwidth]{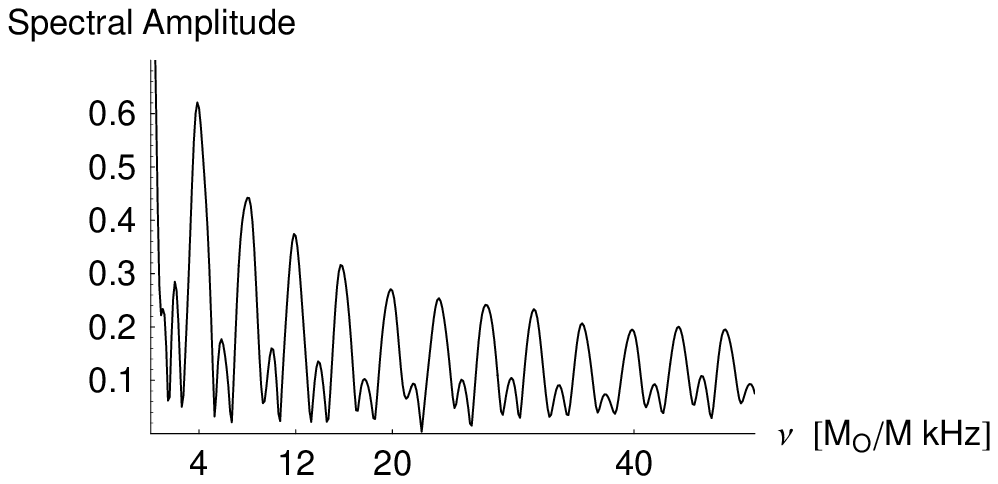}
\caption{\emph{Left}: A typical chirp: light curve produced by a small ($r=0.01GM/c^2$) solid body
on its way to the black hole following a parabolic orbit with angular momentum $l/m=3.999998 G M/c$ which makes three turns before
crossing the horizon. In this example the body's velocity, when far
from the black hole, subtends an angle $24^o$ with the line of sight and the observer is located $5^o$ above the orbital plane. 
The sharp peaks in observed
luminosity are the effect of gravitational lensing. The wide bumps following those peaks are due to aberration of light
and Doppler shift. 
\emph{Right}: The Fourier spectrum of the chirp on the left. Note that the frequency scale is
in kHz, divided by the mass of the black hole in solar units; for example,
a signal from a 10 solar mass black hole would have the lowest peak frequency
at $\nu\sim 400\textrm{Hz}$. This frequency corresponds to the orbital period on the last circular
orbit about the black hole.}
\label{kos:fig1}
\end{figure}
\begin{figure}
\centering
\includegraphics[width=0.49\textwidth]{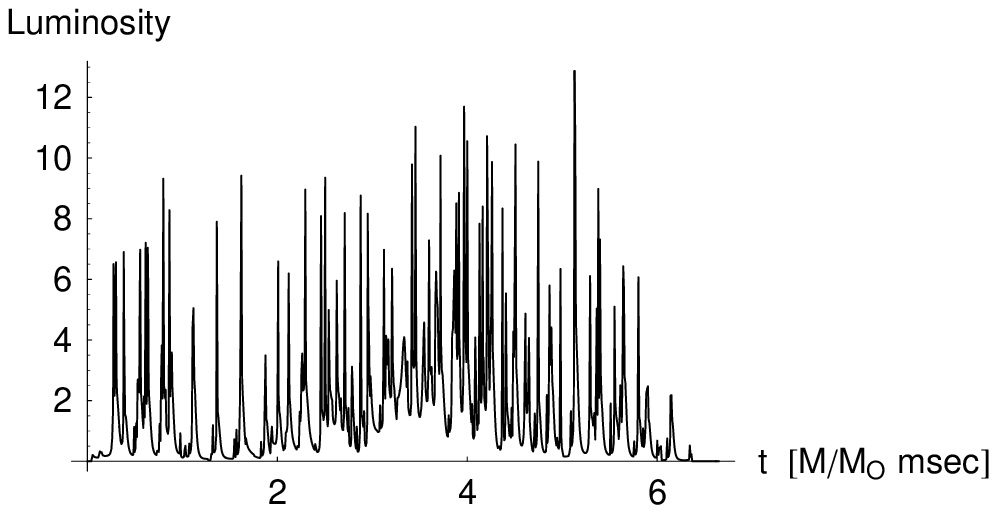}
\includegraphics[width=0.49\textwidth]{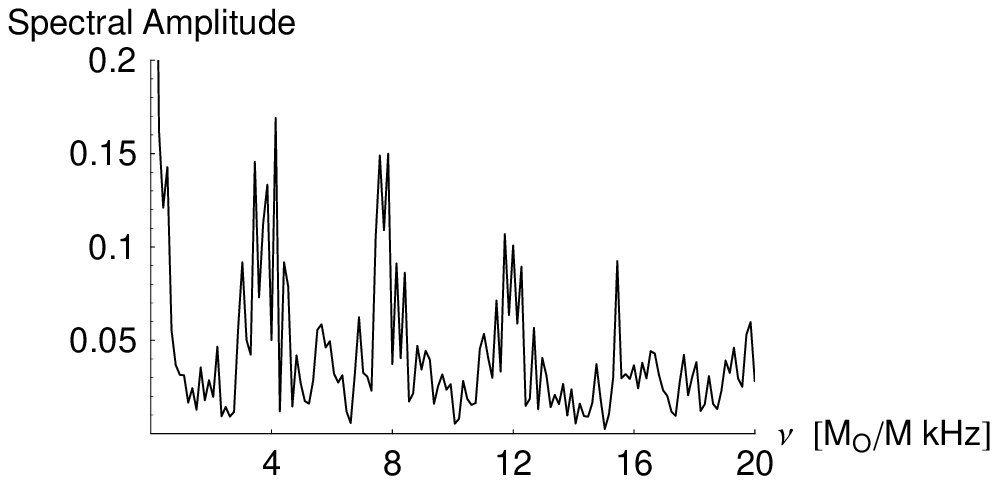}
\caption{A light curve (left) and the Fourier spectrum (right) of a superposition of 24 chirps produced by
24  particles orbiting on $l/m=3.999998 G M/c$ parabolic orbits distributed randomly in time and with respect
to the angle of the line of nodes. The observer is located $5^o$ above the orbital plane.}
\label{kos:fig2}
\end{figure}
\begin{figure}
\centering
\includegraphics[width=0.49\textwidth]{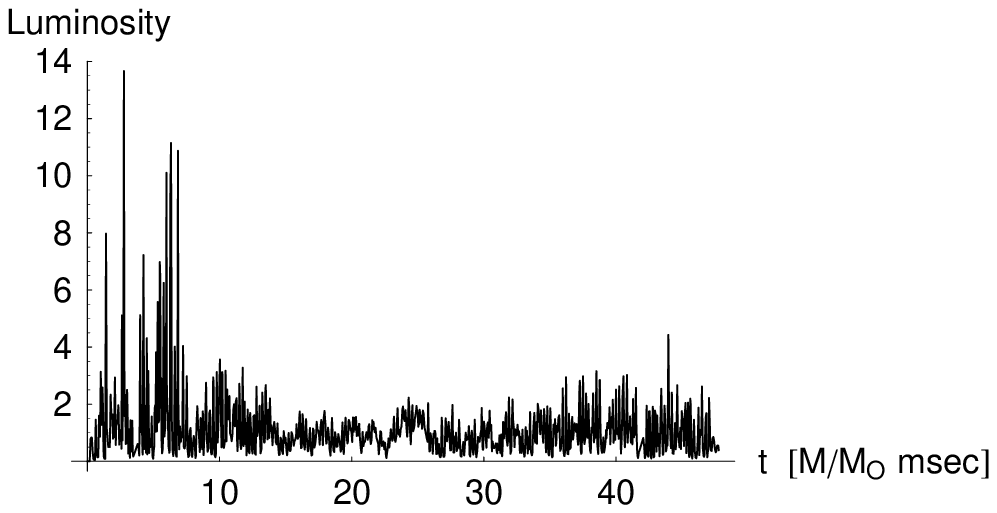}
\includegraphics[width=0.49\textwidth]{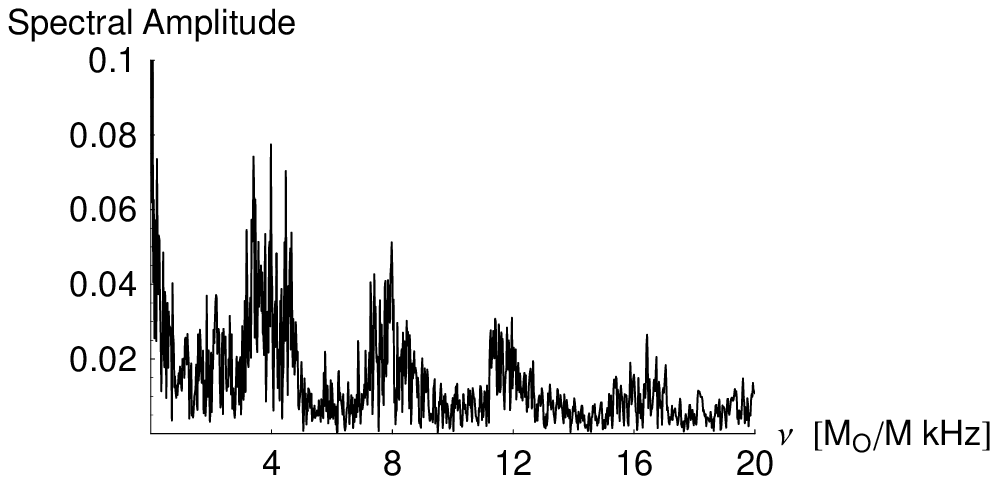}
\caption{Left: A signal produced by 178 objects falling along $l/m=3.999998 G M/c$ parabolic orbits
onto the black hole from random directions
and at random times. Right: Fourier spectrum of this signal. The broad peaks
appear at harmonics of the frequency corresponding
to the period of the last circular
orbit, but their amplitudes are again modulated
by random factors determined by
the relative timing of contributing chirps.
Another random series of chirps would
produce a similar, but not the same power
spectrum. 
}
\label{kos:fig3}
\end{figure}
\section{Conclusions}
Accretion of solid asteroid-like material onto a black hole can produce signals with signature
similar to that of QPOs. Small solid particles are the only form of matter that can
survive huge tidal forces deep in the gravitational field of a stellar mass black hole. They
can enter the last circular orbit almost intact and transform the mounting tidal
force into internal energy. Since the theoretical upper limit for this energy increase is $\sim 0.1
\textrm{mc}^2$, it seems plausible that the body heats to hard X-ray temperatures when making a few
last turns before plunging behind the horizon. Such a signal is modulated by Doppler shift,
aberration of light and gravitational lensing and produces a characteristic chirp. 
In case of Schwarzschild black hole, all signals have the same characteristic frequency $1/50 [c^3/GM]$, but the higher frequencies depend
somewhat on the orientation of the orbit with respect to the observer. Many objects produce
a signal which is very similar to that of a QPO \cite{kos:2005ApJ...628L..53I,kos:2005ApJ...623..383H}.
%
%
%
%

%
%
%
\printindex
\end{document}